\documentclass[twocolumn,showpacs,preprintnumbers,prl,nofootinbib]{revtex4}
\usepackage{amsfonts}
\usepackage{amsmath}
\usepackage{graphicx}
\usepackage{dcolumn}
\usepackage{bm}
\usepackage{latexsym}

\begin{document}

\title{Position Measurements Obeying Momentum Conservation}
\author{Paul Busch}
\email{paul.busch@york.ac.uk}
\author{Leon Loveridge}
\email{ldl500@york.ac.uk}
\affiliation{Mathematical Physics Section, Department of Mathematics, University of York,
York, UK}
\date{\today}

\begin{abstract}
\noindent We present a hitherto unknown fundamental limitation to a basic measurement: that of the position of a 
quantum object when the total momentum of the object and apparatus is conserved. This result extends the 
famous Wigner-Araki-Yanase (WAY) theorem, and shows that accurate position measurements are only 
practically feasible if there is a large momentum uncertainty in the apparatus.
\end{abstract}

\maketitle

\noindent {\bf 1. Introduction.}
The extent to which the elements of the quantum mechanical
formalism relate to physically measurable quantities has been the subject
of many investigations in the history of quantum mechanics.
It is well known, for example, that not all self-adjoint
operators represent observables in the presence of superselection rules. 
Wigner \cite{wigner} showed that a different type of measurement limitation arises due
to conservation laws for quantities that are additive over the
system plus apparatus. Specifically he, and subsequently Araki and
Yanase \cite{way} proved that  
a discrete self-adjoint operator not commuting with such a conserved
quantity does not admit perfectly accurate and repeatable measurements.
The original proofs of the WAY theorem are restricted to cases where the object part of the 
conserved quantity is bounded. If  that quantity is assumed to be discrete,
the second, positive part of the WAY theorem asserts that a repeatable
measurement can be approximately realized, but this comes at a price: high accuracy
requires a large size (suitably defined) of the apparatus  \cite{way,yan}. 

The most comprehensive extensions of the WAY theorem obtained so far \cite{SteinShim,Ghirardi}
do not encompass more general cases  including continuous-spectrum and unbounded 
observables. In fact, it is a fundamental result established by Ozawa \cite{repeatable}  that 
continuous observables do not admit any repeatable measurements, irrespective of whether there 
are additive conserved quantities. 

Nevertheless, our analysis of a model presented by Ozawa \cite{Ozawa} in this journal leads us to 
conclude that WAY-type limitations do exist for measurements of continuous quantities, contrary to 
the view expressed there. We show for the prototypical example of position measurements obeying 
momentum conservation that the accuracy and approximate repeatability of such measurements 
are limited by the  finite size of the apparatus if it is assumed that the pointer 
observable commutes with the momentum. This condition, which following Ozawa \cite{ozawacon} we call the
 \emph{Yanase condition}, is certainly significant but often neglected: 
In order to secure  reproducible
measurement records, it is necessary that the pointer observable itself can be measured repeatably and 
accurately. Insofar as the WAY theorem applies to the pointer observable being measured, this may only 
be achieved if that observable commutes with the conserved quantity.

We also consider an alternative model which shows, perhaps surprisingly, that if one relinquishes 
the Yanase condition, position measurements obeying momentum conservation may be
possible with arbitrary accuracy and good repeatability properties, without any constraint on the 
size of the apparatus. This stands 
in contrast to the discrete-bounded case where a measurement of a quantity not commuting with an 
additive conserved quantity can neither be repeatable nor satisfy the Yanase condition 
\cite{LoveridgeBusch}. We also provide a general, model-independent argument corroborating 
these findings.

A thorough understanding of such quantum limitations to measurements is crucial; 
from a foundational perspective it provides a more complete description of 
physical reality as it manifests itself through observation, and from a pragmatic 
viewpoint it delineates the possible fundamental obstacles that must be accounted for 
in technological applications.
Ozawa and coworkers \cite{conquant} have demonstrated a limitation to the realizability 
of quantum logic gates insofar 
as the observables involved are subject to the WAY theorem. 
Similarly it must now be expected that  operations for continuous-variable
quantum information processing tasks are only realizable to a limited accuracy in the
presence of an additive conservation law, given that there will typically be a need to limit the size of 
the component systems. For accurate position measurements subject to a WAY-type
limitation, a large momentum spread---and thus kinetic energy---is required in the apparatus, which 
conflicts with the low temperatures necessary for the control of a quantum system.

In the models discussed below, the system  and the apparatus
are  particles in one space dimension, represented by the Hilbert space of
square-integrable functions on $\mathbb{R}$. 
We will work in units where $\hbar =1$. \\

\noindent{\bf 2. Ozawa's  model.}
In \cite{Ozawa}, Ozawa claimed that there is no WAY-type limitation to
position measurements. 
 He introduced a model 
involving four particles with position operators 
$Q,Q_{\mathcal{A}},Q_{\mathcal{B}},Q_{\mathcal{C}}$. 
 The interaction Hamiltonian
is translation invariant and thus conserves total momentum; the resulting unitary time evolution
operator for a time interval $\tau$ is  
\begin{equation}
U=\exp \left[ -i\frac{K}{2} \tau (Q-Q_{\mathcal{A}})(Q_{\mathcal{B}}-Q_{\mathcal{C}})\right] .
\end{equation}
$U$ acts on $\mathcal{H}_{total}:=\mathcal{H}\otimes \mathcal{H}_{\mathcal{
A}}\otimes \mathcal{H}_{\mathcal{B}}\otimes \mathcal{H}_{\mathcal{C}}$ and
we adopt the obvious shorthand (e.g., $Q=Q\otimes \boldsymbol{1}_{\mathcal{A}
}\otimes \boldsymbol{1}_{\mathcal{B}}\otimes \boldsymbol{1}_{\mathcal{C}}$)
for simplicity of notation. The constant $K$ describes the 
coupling strength; we will use the abbreviation $K\tau =\lambda$.

The aim is to use this interaction to measure a particle's position, $Q$, by
transcribing the $Q$-distribution to a pointer observable $Z$ 
on an apparatus that is accessible to an experimenter. Here the pointer is taken to be the relative
momentum $Z=P_{\mathcal{C}}-P_{\mathcal{B}}$. With this choice, particle $\mathcal A$  
appears as an auxiliary ``reference''  system by which information about $Q$ can be recovered. 
It is also clear that $[Z,P_{total}]=0$ where $P_{total}$ is sum of the momenta of the system and 
apparatus. Thus the Yanase condition is satisfied in this model.

Ozawa chooses $K\tau=1$ (which  makes position and momentum dimensionless)
 and the  initial apparatus state 
 $\xi =|Q_{\mathcal{A}}=\overline{y}\rangle \otimes $ $|P_{\mathcal{C}}-P_{\mathcal{B}}=
\overline{y}\rangle $ for $\overline{y}$ constant. He omits the state
representing the final degree of freedom pertaining to $P_{B}+P_{C}$, which does not alter the outcome. 
By the uncertainty relation,
this choice of (unnormalizable) initial state $\xi$ cannot have finite momentum spread.

The observable-to-be-measured $Q$ is preserved by the interaction: $Q=Q(\tau )$. 
The characteristic function, which arises as the Fourier transform of the joint probability density of   
 $Q=Q(\tau )$ with the time-evolved pointer observable 
 $Z(\tau )=(P_{\mathcal{C}}-P_{\mathcal{B}})+(Q-Q_{%
\mathcal{A}})$, is given by the expression 
$\left\langle \varphi \otimes \xi |\exp({i(\mu Q(\tau)+\mu ^{\prime }Z(\tau ))})\varphi
\otimes \xi \right\rangle$.  
Ozawa gives this in integral form as 
\begin{equation}
\iint e^{i(\mu x+\mu ^{\prime }z)}
\left\vert \varphi (z)\right\vert^{2}\delta (x-z)dxdz,  \label{oz. int}
\end{equation}
where $z$ denotes a spectral value of $Z$
and $\varphi $ is the preparation of the system. However, this follows only by ignoring the
two-fold infinity generated by the term 
$\left\langle \overline{y}|\overline{y}%
\right\rangle \left\langle \overline{y}|\overline{y}\right\rangle $ that
would appear in the original expression for the characteristic function. 
Thus the distribution $\left\vert \varphi (z)\right\vert^{2}\delta (x-z)dx dz$ following from (\ref{oz. int}) is not the
the joint distribution of $Q(\tau)$ and $Z(\tau)$, and hence it does not follow that this model realizes an
accurate and repeatable measurement of position. This conclusion is in line with Ozawa's
result  that continuous observables do not admit repeatable measurements \cite{repeatable}.

We shall now  calculate the relevant measurement probabilities
directly in the Schr\"{o}dinger picture, using normalizable
states only  \cite{KK}. 
It follows that  the measurement accuracy---and degree of repeatability (see Sec.~4)---are limited
by the ``size" of the apparatus, in close analogy to what we referred to as the positive part of
the WAY theorem in the case of discrete quantities. Here we use the position and momentum representations for the initial (product) state, 
$\Psi _{0}(x,y,u,v)=\varphi (x)\Phi _{1}(y)\Phi _{2}(u)\phi(v)$ 
with $u$ and $v$ denoting spectral values of $P_{\mathcal{C}}-P_{\mathcal{B}}$ and 
$P_{\mathcal{B}}+P_{\mathcal{C}}$, respectively. After a time $\tau$ (which we will also write as $\lambda/K$), the state has evolved into
\begin{equation}
\Psi_\tau(x,y,u,v)=\varphi(x)\Phi_1(y)\Phi_2\bigl(u+\frac 12\lambda(x-y)\bigr)\phi(v).
\end{equation}

The probability density for $u$ is obtained as a marginal from the joint density 
for the time-evolved state $\Psi _{\tau}$;
\begin{equation}
p_{\Psi_\tau} (u)=
\iiint
\bigl\vert  \Psi _{\tau}(x,y,u,v)\bigr\vert^{2}dxdydv.
\end{equation}
The probability for the pointer to assume a value in a set $S$  is:
\begin{equation} \label{joint}
\begin{split}
\mathsf{P}_{\Psi_\tau} &(u \in S)  =\int_ { S} du \int dx\int dy
\left\vert \varphi(x)\right\vert ^{2} \\
&  \times \left\vert \Phi _{1}(y)\right\vert ^{2}\bigl\vert \Phi _{2}\big(u+
\frac{1}{2}\lambda (x-y)\big)\bigr\vert ^{2}\int dv|\phi(v)|^2.
\end{split}
\end{equation}
We introduce a \emph{scaling function} $f:\mathbb{R}\to\mathbb{R}$ 
to allow for the measured observable and the pointer observable to have different scales. 
With $f(u)=-(2/\lambda)u$ and putting $S= f^{-1}(X)=-(\lambda /2)X$ (the set of all $u$
with $f(u)\in X$), 
the right hand side of \eqref{joint} can be written as:
\begin{align}\label{eqn:probQ}
&\int dx\left\vert \varphi(x)\right\vert ^{2} \chi_{X} \star e^{(\lambda)}(x)\nonumber\\
&\quad=\int_X \int dx'\left\vert \varphi(x+x')\right\vert ^{2} e^{(\lambda)}(x')\equiv \mathsf{P}_\varphi(x\in X).
\end{align}
with $\star$ denoting the \emph{convolution} and $\chi_X$ the set indicator function.
The function $e^{(\lambda)}$ is a density and takes the form 
$e^{(\lambda )}(x)=(\vert \Phi_{1}\vert ^{2}\star \vert \Phi _{2}^{(\lambda )}\vert^{2})(x)$,
where $\Phi _{2}^{(\lambda )}(s)=\sqrt{\lambda}\Phi_2(\lambda s)$.
 This density function $e^{ (\lambda)}$ represents the inaccuracy of the measurement \cite{POVM}, 
in the sense that the actual probability density appearing in (\ref{eqn:probQ}) is a smearing
of the ideal position probability density $|\varphi(x)|^2$; we see that
  the narrower the width 
 of $e^{ (\lambda)}$, the more accurate the measurement. 
In the extreme case that $e^{ (\lambda)}$ tends to a delta-function, the probabilities 
(\ref{eqn:probQ}) become those of an accurate position measurement.

We compute
$\mathrm{Var}(e^{(\lambda )})=\mathrm{Var}\left\vert \Phi _{1}\right\vert
^{2}+\frac{4}{\lambda ^{2}}\mathrm{Var}\left\vert \Phi _{2}\right\vert ^{2}$. 
Thus the variance of $e^{(\lambda)}$ does not vanish in the limit $\lambda\to\infty$ but is given by
the variance of the $Q_{\mathcal{A}}$ distribution in the 
``reference system" state $\Phi _{1}$; by virtue of the
uncertainty relation for $Q_{\mathcal{A}}$ and $P_{\mathcal{A}}$, this can only be
made small at the expense of making the width of the $P_{\mathcal{A}}$
distribution large. We see that in order to recover accurate information
about the particle's position $Q$, it is the reference position $Q_{\mathcal{A}}$ that needs to be highly
localized, independently of the momentum spread of the pointer.

In accordance with the findings of Yanase \cite{yan} for the case where the
object part of the conserved quantity was bounded and discrete, we see here
that the size of the apparatus limits the position measurement accuracy.

A more useful measure of inaccuracy than the variance of a distribution $e$ is given by
the \emph{overall width} $W(e;1-\varepsilon )$ of $e$ at confidence level 
$1-\varepsilon $, defined as the smallest possible size of a suitably located
interval $J$ such that the probability $\int_{J}e(q)dq\geq 1-\varepsilon $.
In contrast to the variance, the overall width is finite whenever $\varepsilon >0$. 
 
It is straightforward to show that the
overall width of a convolution of two probability distributions is bounded 
below by the width of the largest. In the case of the Ozawa model, we thus see that
the overall width of $e^{(\lambda )}$ is always bounded below by the overall width of the 
distribution $|\Phi_1|^2$, which is independent of $\lambda$. This generalizes the above argument
which used variances.\\

\noindent{\bf 3. An alternative model.}
Next we revisit a position measurement model 
\cite[Sec.~IV.3.3]{OQP} that violates the Yanase condition. Momentum conservation is implemented
via the unitary coupling
\begin{equation}
U=\exp \left[ -i\frac{\lambda }{2}\bigl((Q-Q_{\mathcal{A}})P_{\mathcal{A}}
+P_{\mathcal{A}}(Q-Q_{\mathcal{A}})\bigr)\right],
\end{equation} 
which acts on $\mathcal{H}\otimes \mathcal{H}_{\mathcal{A}}$. As before, $\lambda$ is a shorthand
for $K\tau$ where $K$ is the coupling strength and $\tau$ the duration of the interaction. Here $\lambda$ is
naturally dimensionless.
The pointer observable is $Q_{\mathcal{A}}$, which of course does not commute with the total momentum.

We can again extract the probability density for the pointer after time $\tau$, with $\Psi_\tau=U(\varphi\otimes\phi)$:
\begin{equation}
p_{\Psi_\tau} (y)=
\int \left\vert  \Psi _{\tau}(x,y)\right\vert^{2}dx.
\end{equation}
The form of the final state $\Psi _{\tau}(x,y)$ gives the pointer probabilities
\begin{equation}
\begin{split}
\mathsf{P}_{\Psi_\tau} (y \in f^{-1}(X)) & = \int_{ f^{-1}(X)} dy \int dx \left\vert \varphi
(x)\right\vert ^{2}   \\
& \times e^{\lambda }\left\vert \phi (y e^{\lambda}-x(e^{\lambda }-1)\right\vert ^{2},
\end{split}
\end{equation}
which, with $f^{-1}(X):=(1-e^{- \lambda})X$, we write in the form
\begin{equation}
\begin{split}
\int dx \left\vert \varphi
(x)\right\vert ^{2} \chi _{X} \star e^{(\lambda)}(x) \equiv \mathsf{P}_{\varphi} (x \in X).
\end{split}
\end{equation}
The probability density $e=e^{(\lambda )}$ now takes the form 
$e^{(\lambda )}(x)=(e^{\lambda }-1)\left\vert \phi (-x(e^{\lambda}-1))\right\vert ^{2}.
$ 
 The scaling behavior is thus \emph{exponential} in $\lambda$; the inaccuracy width scales with $e^{-{\lambda }}$ 
 and an arbitrarily accurate measurement of $Q$ is feasible without any constraint on the size of the apparatus. \\

\noindent{\bf 4. Repeatability.} 
It is worth elucidating further the differences between the two models studied here. The first, which satisfied the 
Yanase condition, displayed limitations to the accuracy of a position measurement that could be overcome only 
by allowing the reference system to have large momentum. The second, which manifestly violated the Yanase 
condition, imposed no such constraint and arbitrary accuracy could be achieved by a  tuning of the interaction 
strength. However, as in the original work \cite{wigner,way}, it is not only the measurement accuracy that
plays a prominent role, but also the repeatability properties, which we discuss now.

We shall confine the probe's initial state wavefunctions to a bounded subset of the real line. This is not an overly
stringent requirement from a physical perspective. In the Ozawa model this simply amounts to the initial state functions $\Phi_1(y)$ 
and $\Phi_2(u)$ having finite extent in the relevant variables; in the second model it means that the probe state function $\phi(y)$
is concentrated in a finite interval. Thus we can think of the density $e^{(\lambda)}$ as being concentrated on the interval $[-d,d]$
in either model.

One way of quantifying the degree of approximate repeatability \cite{Davies,BL} in the case of a position measurement is as follows: A measurement is said to be approximately, or $\delta$-repeatable
if given an outcome in a set $X$, the outcome of an immediate subsequent control measurement 
will be found, with probability 1, in a suitably enlarged set $X_{\delta}$ 
(where $X_{\delta}$ is the set of points not more than a distance $\delta > 0$ away from $X$). This 
can be written symbolically as a conditional probability of finding the particle's position $x\in X_\delta$ given that the pointer was found to have 
a value $u\in f^{-1}(X)$:
\begin{equation} \label{eqn:delt}
\mathsf{P}_{\Psi_\tau}\bigl(x \in X_{\delta}|u \in f^{-1}(X)\bigr)=1
\end{equation}
for all sets $X$. Considering the control measurement to be accurate, for this to be satisfied in the Ozawa model we must have
$\chi_{X}\star e^{(\lambda)}(x)=0$ whenever $x$ is outside $X_ \delta$, and this follows if $\delta \geq d$. 
If the initial apparatus states $\Phi_1$ and $\Phi_2$ are concentrated on intervals $[- \ell,\ell]$ and $[-m,m]$ respectively,
we have that $d= \ell + m/ \lambda$. Therefore even as the coupling strength $\lambda$ becomes large, $\delta$ is bounded below by the
width of the reference system state $\Phi_1$, and in order to recover good repeatability properties (i.e. a small $\delta$),
the state $\Phi_1$ must carry a large spread of momentum.

In the alternative model we see similar behavior, with a fundamental difference; we again have that
$\delta \geq d$ enables  approximate repeatability in the sense of \eqref{eqn:delt}. However, in contrast to the Ozawa model, simply 
letting $\lambda$ be large allows for arbitrarily good repeatability; if $\phi$ is concentrated on $[-n,n]$, then $d=n/(e^{\lambda}-1)$.

Thus under violation of
the Yanase condition, arbitrarily accurate and repeatable information
transfer from the system to a quantum probe is feasible without 
any size constraint ($n$ can be arbitrarily large, allowing the spread of the probe momentum to be small).\\

\noindent{\bf 5. General argument.}
Finally we adapt an approach due to Ozawa \cite{ozawacon} to obtain a
generic, model-independent trade-off between the qualities of accuracy and
repeatability on one hand and the necessary ``size" of the apparatus on the 
other. The
noise operator $N$ is defined as $N:=Z(\tau )-Q$, where $Z(\tau )$
represents the Heisenberg-evolved pointer observable after the interaction
period $\tau $. One then defines the \textit{noise }$\epsilon (\varphi
)^{2}:=\left\langle \varphi \otimes \phi |N^{2}\varphi \otimes \phi
\right\rangle \equiv \langle N^{2}\rangle $. Clearly $\epsilon (\varphi
)^{2}\geq (\Delta N)^{2}$. For a measurement scheme to represent an approximation
to a position measurement, it is reasonable to require that the noise is finite across 
all input object states. Thus the supremum  $\epsilon :=\sup \epsilon (\varphi )$ should
be finite and would then give a global measure of {\em error}.
 The uncertainty relation then gives 
\begin{equation}\label{eps}
\epsilon^2\ge \epsilon (\varphi )^{2}\geq \frac{1}{4}\frac{\left\vert \left\langle \left[
Z(\tau )-Q,P+P_{\mathcal{A}}\right] \right\rangle \right\vert ^{2}}{(\Delta
P_{total})^{2}},
\end{equation}%
where $(\Delta P_{total})^{2}=(\Delta _{\varphi }P)^{2}+(\Delta _{\phi }P_{\mathcal{A}})^{2}$. 
This inequality entails a measurement limitation whenever
the right hand side is nonzero for some object states. It is also evident that if the numerator is nonzero, the only
way of making this lower bound to the error small independently of the object properties
is by making the momentum variance $(\Delta _{\phi }P_{\mathcal{A}})^{2}$ of the apparatus large.

The vanishing of the numerator for all object states  $\varphi$ follows when the commutator
is zero, which happens just when the pointer at time 0 satisfies $[Z,P_{\mathcal A}]=i$.
This is the case in the second model discussed above where a WAY-type limitation was found 
to be absent. 

If the Yanase condition is stipulated, one obtains
$\left[ Z(\tau )-Q,P+P_{\mathcal{A}}\right] =i$, and (\ref{eps}) yields
\begin{equation}
\epsilon ^{2}\geq 
 [2\Delta _{\phi }P_{\mathcal{A}}]^{-2}.
\end{equation}%
This bound only allows for an increase in
accuracy when $(\Delta_\phi P_{\mathcal{A}})^{2}$ is large,
thus establishing necessity of the large apparatus size for good measurements.

An attempt at capturing (approximate) repeatability in the generic case follows from
considering the quantity $\mu (\varphi )^{2}:=\langle \varphi \otimes \phi
|(Q(\tau )-Z(\tau ))^{2}\varphi \otimes \phi \rangle $; intuitively if this
expectation is small, then the difference between the measured observable
and the time-evolved system observable is small, and hence the measurement
should display some level of repeatability.\ An argument analogous to that
above gives, for $\mu ^{2}:=\sup \mu (\varphi )^{2}$ 
\begin{equation}
\mu ^{2}\geq   
  [2\Delta _{\phi }P_{\mathcal{A}}]^{-2}.  \label{app_rep}
\end{equation}%
This provides an indication that under the Yanase condition, good repeatability is  achieved, again,
only when there is a large momentum variance in the apparatus. It remains to be shown
that these conclusions persist when more operationally significant measures of
inaccuracy and repeatability are used, such as those in \cite{teiko}. For example, a new
measure of repeatability may be formulated via the \textit{repeatability
width}, defined as the smallest $\delta $ such that a repeatability condition like
(\ref{eqn:delt}) is
satisfied, possibly only up to probabilities no less than a threshold $1-\varepsilon$.

In conclusion, evidence for a WAY-type theorem for continuous unbounded quantities 
has been provided through two models of momentum-conserving position measurements and
two model-independent inequalities. The analysis entails also that 
no such limitation arises if only \emph{relative} 
distances are measured, that is the distance between the object and the ``reference system'',
which is provided by the measuring apparatus. When this is incorporated into
the quantum description, the conservation law can be manifestly satisfied for the combined object-apparatus
system, with the measured observable as the relative position. In this case, the Yanase condition
must be satisfied for good accuracy to be achieved.
This points to a possible connection, hinted at by Aharonov and Rohrlich \cite{AR}, with the theory of 
superselection rules and quantum frames of reference,
a subject of renewed interest in the past decade \cite{qfr}, which seems to deserve further systematic study. 

The authors would like to thank an anonymous referee for valuable comments.


\begin{thebibliography}{99}

\bibitem{wigner} E.~Wigner,  
\emph{Z.~Phys}\textit{.} \textbf{133}, 101 (1952). English translation available at arXiv:1012.4372.

\bibitem{way} H.~Araki, M.M.~Yanase,  
\emph{Phys.~Rev}\textit{.} \textbf{120}, 622 (1960).

\bibitem{yan} M.M.~Yanase, \emph{Phys.~Rev.}~\textbf{123}, 666 (1961).

\bibitem{SteinShim} H.~Stein, A.~Shimony, in {\em Foundations of Quantum Mechanics},
B.~d'Espagnat (ed.), Academic Press, New York, 1971, p.~56.

\bibitem{Ghirardi} G.C.~Ghirardi, A.~Rimini, T.~Weber, {\em J.~Math.~Phys.}~{\bf 24},
2454 (1983).

\bibitem{repeatable} M.~Ozawa,  
\emph{J.~Math.~Phys}\textit{.}~\textbf{25}, 79 (1984).

\bibitem{Ozawa} M.~Ozawa,  
\emph{Phys.~Rev.~Lett}\textit{.}~\textbf{67}, 1956 (1991).

\bibitem{ozawacon} M. Ozawa, 
\emph{Phys.~Rev.~Lett}\textit{.}~\textbf{88}, 050402 (2002).



\bibitem{LoveridgeBusch} L.~Loveridge, P.~Busch,
 \emph{Eur.~Phys.~J.~D},
in press (2011); arXiv:1012.4362 [quant-ph].

\bibitem{conquant} M.~Ozawa, 
\emph{Phys.~Rev.~Lett}\textit{.}~\textbf{89}, 057902 (2002); 
T.~Karasawa, J.~Gea-Banacloche, M.~Ozawa, {\em J.~Phys.~A: Math.~Theor.}~{\bf 42}, 225303 (2009).

\bibitem{KK}  K.~Kakazu \emph{et al}
(\emph{Phys.~Lett.~A}~{\bf 173}, 92 (1993)) do address the issue of unnormalizable pointer 
states and agree with Ozawa's conclusion, but they do not consider 
any  trade-off between apparatus size and measurement accuracy.

\bibitem{POVM} An effective description of the approximate measurements being discussed here is afforded by the language
of \emph{positive operator valued measures}.

\bibitem{OQP} P.~Busch, M.~Grabowski, P.~Lahti, \emph{Operational Quantum
Physics}, Springer, Berlin, 1995/1997.


\bibitem{Davies} E.B.~Davies, 
\emph{J.~Func.~An.}~\textbf{6}, 318-346 (1970).

\bibitem{BL} P.~Busch, P.~Lahti, 
\emph{Ann.~Physik (Leipzig)} \textbf{47}, 349 (1990).


\bibitem{teiko} C.~Carmeli, T.~Heinonen, A.~Toigo, \emph{J.~Phys.~A: Math.~Theor.} \textbf{40}, 1303 (2007).

\bibitem{AR} Y.~Aharonov, D.~Rohrlich, \emph{Quantum Paradoxes -- Quantum Theory for the Perplexed}, 
Wiley-VCH Verlag, Weinheim, Germany, 2005, Chapter 11.

\bibitem{qfr} S.D.~Bartlett, T.~Rudolph, R.W.~Spekkens, {\em Rev.~Mod.~Phys.}~{\bf 79}, 555 (2007).

\end{thebibliography}
\end{document}